\documentclass[12pt]{article}
\usepackage{epsf}
\def\be{\begin{equation}}
\def\ee{\end{equation}}
\def\bea{\begin{eqnarray}}
\def\eea{\end{eqnarray}}

\begin{document}
\begin{titlepage}
\begin{center}
{\Large \bf William I. Fine Theoretical Physics Institute \\
University of Minnesota \\}
\end{center}
\vspace{0.2in}
\begin{flushright}
FTPI-MINN-15/30 \\
UMN-TH-3441/15 \\
June 2015 \\
\end{flushright}
\vspace{0.3in}
\begin{center}
{\Large \bf Condensed state of heavy vectorlike neutrinos
\\}
\vspace{0.2in}
{\bf M.B. Voloshin  \\ }
William I. Fine Theoretical Physics Institute, University of
Minnesota,\\ Minneapolis, MN 55455, USA \\
School of Physics and Astronomy, University of Minnesota, Minneapolis, MN 55455, USA \\ and \\
Institute of Theoretical and Experimental Physics, Moscow, 117218, Russia
\\[0.2in]

\end{center}

\vspace{0.2in}

\begin{abstract}
Some extensions of the Standard Model suggest existence of vectorlike fermions whose mass $M$ can be by up to several orders of magnitude larger than the electroweak scale $v$, and with the Higgs mechanism providing only a small part $m$ of this mass that is a fraction of $v$. It is shown that in a certain range of the parameters $m$ and $M$  there exist stable globules  of degenerate fermionic matter made from the lighter of the neutral vectorlike leptons (superheavy neutrinos) with a number density much larger than $m_h \, v^2$, with $m_h$ being the Higgs boson mass. The stability is due to the deformation of the Higgs field inside the globule to values greatly exceeding $v$. The globules are absolutely stable if the lighter of the vectorlike neutrinos is stable and can be (a part of) the dark matter. 
\end{abstract}
\end{titlepage}

An addition of new heavy vectorlike families of quarks and/or leptons to the Standard Model generally does not lead to contradictions with the existing constraints from experiment.  The vectorlike fermions can have a mass term that is a singlet under the electroweak (EW) gauge group $SU(2)$. Thus such fermions are allowed to acquire `a mass of their own', not constrained by the condition of absence of a strong Yukawa coupling to the Higgs boson, so that their mass scale $M$ can be much larger  than the EW scale $v = (G_F \, \sqrt{2})^{-1/2} \approx 250$\,GeV. A possible weak Yukawa interaction with the Higgs field provides a relatively small splitting $m$ between the masses of these fermions. (In particular, increasing the Yukawa coupling above values corresponding to $m \sim 100\,$GeV can create problems with the Higgs vacuum stability~\cite{dveegis}.)  Models containing vectorlike fermions were considered in the literature on numerous occasions either in the context of purely theoretical constructions such as a unification of gauge couplings~\cite{bp,Dermisek12,Dermisek13}, or in connection with their specific phenomenological signature: the muon $g-2$ anomaly~\cite{cm,krss,dr}, the  precision electroweak measurements~\cite{ctw} and in the Higgs boson decays~\cite{df,kpw,clw,alw,jsw,ahbdf,abmf,dlm,mv12}. A comprehensive survey of models with vectorlike fermions can be found in Ref.~\cite{eggw} and a recent analysis of limits on such models from LHC data is presented in Ref.~\cite{dhls}. It can be also noted that if vectorlike fermions carry a conserved global quantum number the lightest of them, most likely electrically neutral superheavy neutrinos, constitute a part or all of the dark matter.

If vectorlike superheavy neutrinos do indeed exist and were at some point present in a significant abundance in the Universe, it makes sense to consider their collective behavior at a finite density. In what follows the energy of the ground state with a uniform number density $n$ of the lightest seperheavy neutrinos is considered in a model with vectorlike leptons. As a result it is found that in a significant part of the parameter space of such models the neutrinos may form globules with the density that at large $M$ scales as $n \propto f \, M^3 m^2/v^2$ with a small factor $f$ being a function of the ratio $m/m_h$, where $m_h \approx 125$\, GeV is the mass of the Higgs boson.  The stability of such object with respect to dissociation into individual fermions arises from a deformation of the Higgs field $h$ by the Yukawa interaction with the superheavy neutrinos to a value greatly exceeding its vacuum average: in the limit of large $M$ the Higgs field inside the globule scales as $h \propto g \, M m^{1/3}/v^{1/3}$ with a numerically small factor $g$ that depends on $m/m_h$. The enhancement of the Higgs field inside the condensate lowers the effective mass of the neutrino and provides a reduction of the overall mass of the globule below the sum of the masses of the constituent fermions in vacuum. A detailed consideration of possible dynamics of the process of formation of the globules is beyond the scope of the present study. It will just be argued that even if a sufficient density of the neutrinos is created, there is still an energy barrier in the deformation of the Higgs field from its vacuum value $v$. Thus the formation of the discussed condensed objects in the presently existing vacuum is very improbable. However, if, in the course of cosmological evolution, the Higgs field in the vacuum had evolved from larger values, the globules could well be left over until the present epoch. 

For concreteness the treatment here is limited to a minimal model with one generation of vectorlike leptons with a conserved fermion number. The model thus contains two EW $SU(2)$ doublets and at least two singlets which can be labeled as being of opposite chirality:
\be
L= \left ( \begin{array}{c}
             N_1 \\
             E 
           \end{array} \right )_L ~,~~~
R= \left ( \begin{array}{c}
             N_1 \\
             E 
           \end{array} \right )_R ~,~~~ N_{2L}~,~~~N_{2R} 
\label{ne}
\ee 
with $N_1$ and $N_2$ being electrically neutral and $E$ - charged. Naturally, charged singlets can be added to the model, but  the minimal scheme with only neutral singlets is sufficient for the purpose of the present paper. Moreover, a further restriction of exact $L \leftrightarrow R$ symmetry (the parity conservation in the vectorlike sector) is assumed in the present treatment, so that 
the mass term and the Yukawa interaction with the Higgs doublet $\phi$ is written as
\be
\Delta {\cal L} = - M_1 \,  ( \bar L R) - M_2 \, ( \bar N_{2L} N_{2R} ) - y \, \phi \, \left [ (\bar L N_{2R}) + (\bar R N_{2L}) \right ] + {\rm h.c.}
\label{my}
\ee
where $M_1$ and $M_2$ are the EW singlet mass parameters, and $y$ is the Yukawa coupling. In this scheme the charged fermion acquires mass $M_1$, while the mass term for the two Dirac neutral fermions $N_1$ and $N_2$ has the form
\be
{\cal L}_N  = - M_1 \, (\bar N_1 N_1) - M_2 \, (\bar N_2 N_2) - m \, ( 1+ \chi) \, \left [ (\bar N_1 N_2) + (\bar N_2 N_1) \right]~,
\label{mn}
\ee
where $m = \sqrt{2} \, y v$, and the notation $\chi=h/v$ is introduced in a constant Higgs field, whose neutral component is parametrized in terms of the vacuum mean value $v$ and the deviation $h$ as $\phi^0 = (v+h)/\sqrt{2}$. 
Clearly, the mass eigenvalues for the two neutrinos are
$M_{\pm}= M \pm \sqrt{\Delta^2 + m^2 (1+\chi)^2}$
where $M = (M_1+M_2)/2$ and $\Delta = (M_1-M_2)/2$, and the lighter of the two neutrinos is stable with respect to weak decay within this model. It is for this lightest of the two neutrinos that a state with a finite number density $n$ is considered here.  

Let us start with considering the limit $\Delta=0$, so that the mass of the lighter neutrino in the Higgs vacuum is $M_N = M-m$, and later discuss the case of nonzero $\Delta$. In a degenerate fermion state with number density $n$ the neutrinos fill up the momentum space up to the Fermi momentum $p$, such that $p^3/(3 \pi^2)=n$. Thus the particles are nonrelativistic inasmuch as $n \ll M^3$, which condition, as will be seen further, is satisfied in all the range of $M$ and $m$, where the discussed effect takes place. Allowing for the deviation $h$ of the Higgs field inside the medium to be nonzero, one can write the expression for the difference $\delta \epsilon (n,h)= \epsilon(n,h) - [M-m(h=0)] n$ between the energy density of the medium and the sum of the masses of the neutrinos at the density $n$:
\be
\delta \epsilon(n,h) = {m_h^2 v^2 \over 2} \, \chi^2 \left (1+{\chi \over 2} \right )^2 +  {(3 \pi^2 \, n)^{5/3} \over 10 \pi^2 M}+ {n^2 \over 4 v^2 \, (1+\chi)^2} - m \, \chi \, n~.
\label{de}
\ee
The discussed effect of the neutrino condensation is driven by the last term in this formula describing the shift in the neutrino mass under the shift of the Higgs field $h=\chi v$ which is the only one term that can be negative. The first term on the right hand side describes the energy density of the shifted Higgs field, and the second term is the density of the kinetic energy of the degenerate nonrelativistic neutrinos, where it is assumed that $m (1+ \chi) \ll M$. Finally, the quadratic in the density third term on the right hand side in Eq.(\ref{de}) arises from the self-interaction due to the $Z$ boson force. Indeed, a matter with a finite density of neutrinos has a density of `charge', proportional to $n$, with respect to the $Z$ boson, and the third term describes the positive energy due to the effective Fermi self interaction of a neutral current, where the shift of the Higgs field mean value is also taken into account. Also the coefficient in this term reflects the fact that at $\Delta=0$ the lighter neutrino is an equal mixture of an $SU(2)$ singlet and a doublet, so that the density of the $Z$ charge is generated by only one half of the number density of the particles.

The condensation of neutrinos into globules is possible if $\delta \epsilon$ is negative at some values of $n$ and $\chi$, in other terms if the negative last term in Eq.(\ref{de}) overcomes the sum of the three positive terms. One of the latter terms, the one with the kinetic energy, can be made arbitrarily small by (temporarily) considering the limit $M \to \infty$. In this limit the condition for $\delta \epsilon$ to go negative does not depend on $M$, and it is a straightforward elementary algebra exercise to establish that the minimum of $\delta \epsilon$ is indeed negative if the mass parameter $m$ is sufficiently large:
\be
m > {m_h \over 2 \sqrt{2}} \approx 44\,{\rm GeV}~.
\label{cm}
\ee
Provided that this condition is satisfied, the minimum of $\delta \epsilon$ stays negative at a sufficiently large, but finite, mass $M$. In order to estimate the asymptotic behavior of the density and the Higgs field at large $M$ one can notice that the kinetic energy term does not depend on $\chi$, so that a minimization of the energy density with respect to $\chi$ at a fixed $n$ does not depend on $M$. At $n \gg m_h \, v^2$ one readily finds that at the minimum the value of $\chi$ is also much larger than one and is given by
\be
\chi^3 \approx r  \, {n \over m_h v^2}
\label{chim}
\ee   
with the definition $r =  \sqrt{1+m^2/m_h^2} + m/m_h$. Upon substitution in Eq.(\ref{de}) this gives the dependence of the minimal energy density on $n$ and $M$:
\be
\delta \epsilon_{min} (n) \approx - {(8 \, r \, m/m_h - 2 - r^2) \over r^{2/3}} \, {m_h^2  v^2 \over 8} \, \left ( {n \over m_h v^2} \right )^{4/3} + {(3 \pi^2 \, n)^{5/3} \over 10 \pi^2 M}~.
\label{dem}
\ee
One can readily notice that the condition (\ref{cm}) is exactly the one for the coefficient of the $n^{4/3}$ term to be negative. Once this condition is satisfied the expression in Eq.(\ref{dem}) develops a negative minimum at the density $n$ such that
\be
{n \over m_h v^2}= {(8 \, r \, m/m_h - 2 - r^2)^3 \over 243 \, \pi^4 r^2} \, { m_h \, M^3 \over v^4}~.
\label{nmin}
\ee
Clearly, the proportionality of $n$ to $M^3$ justifies at large $M$  the assumption $n \gg m_h \, v^2$, used in the derivation. On the other hand, Eq.(\ref{nmin}) gives $n \ll M^3$, which justifies the nonrelativistic limit used in the evaluation of the kinetic energy of the neutrinos. Also, according to Eq.(\ref{chim}), the deformation of the Higgs mean value inside the condensate is strong, $\chi \gg 1$, but still $\chi \ll M/m$, so that the shift of the neutrino mass by the Yukawa interaction is small as compared to $M$.

Having established the effect of the condensation of the (superheavy) neutrinos at a large $M$, it is natural to address the question of the minimal value of $M$ at a given $m$, satisfying the condition (\ref{cm}), at which the condensation is possible. Near the critical value of $M$ the value of $\chi$ at the minimum of the energy density is generally not very large, the finite terms in the $\chi$ dependence in Eq.(\ref{de}) can not be neglected, and the minimization is better done numerically. In performing such minimization it is helpful to first minimize the energy density with respect to $n$. Indeed, the expression in Eq.(\ref{de}) is polynomial in $n^{1/3}$ and the minimization in $n$ is equivalent to solving a cubic equation, which can be done analytically. The remaining after that minimization in $\chi$ can be readily done numerically. The result of such numerical evaluation is illustrated in Fig.~1, where the minimal value of $M$, above which the condensation takes place, is shown as a function of $m$. As it follows from the previous discussion the minimal value of $M$ approaches infinity when $m$ goes down to the 
value in the r.h.s. of Eq.(\ref{cm}).

\begin{figure}[ht]
\begin{center}
 \leavevmode
    \epsfxsize=10cm
    \epsfbox{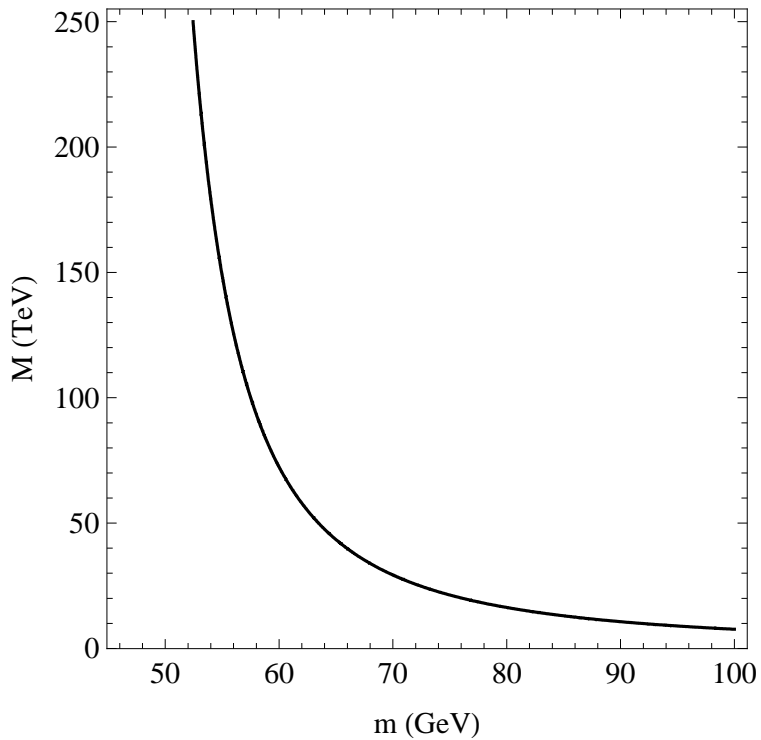}
    \caption{The dependence on the parameter $m$ of the minimal value of the  mass $M$, above which the condensation of heavy vectorlike neutrinos into globules is possible. }
\end{center}
\end{figure} 

It should be also noted that at any given $n$ the energy density $\delta \epsilon$ is always positive at $\chi \ll 1$ which can be readily verified by considering the expansion of the expression in Eq.(\ref{de}) to the second order in $\chi$. Thus a formation of the discussed globules always encounters a barrier before the Higgs field mean value is deformed by at least $h \sim O(v)$.  The scale for the height of barrier is set by Higgs energy density $ \sim m_h^2 v^2$. Therefore in any reasonable situation at present the formation of the globules is essentially impossible. 

The expression in Eq.(\ref{de}) for $\delta \epsilon$ assumes, as previously mentioned, that the mass term (\ref{mn}) contains the degeneracy $M_1=M_2$, so that $\Delta =0$. For a nonzero $\Delta$ one can consider two cases: when $\Delta$ is comparable to $M$~\footnote{The case $\Delta=M$ is the well known `see-saw' mechanism for the masses of the ordinary (chiral) neutrinos.}, and the case $\Delta \ll M$. The first case in fact splits into two sub-cases corresponding to different sign of $\Delta$. If $\Delta < 0$, the lighter neutrino is mostly a component of an $SU(2)$ doublet and the positive contribution to the energy density due to the $Z$ boson field enters with a four times larger coefficient than in in Eq.(\ref{de}). On the other hand, as long as $\Delta \gg m (1+\chi)$, the effect of the mass dependence on the Higgs mean field is approximately equivalent to replacing in Eq.(\ref{de}) the parameter $m$ by $m^2 (1+\chi/2)/\Delta \ll m$, so that the formation of the condensate is impossible since the effective parameter would violate analog of the condition (\ref{cm}). If however $m (1+\chi) \sim \Delta \sim M$, the neutrinos inside the globule have to be relativistic, in which situation the effect of the (weak) dependence of mass on the Higgs field mean value cannot compensate the positive kinetic energy of the particles. The sub-case where $\Delta > 0$ is potentially more interesting, since the lighter neutrino is then mostly a singlet, and the self-repulsion through the $Z$ boson is greatly reduced. In the limit $m (1+ \chi) \ll \Delta$ the difference $\delta \epsilon$ of the energy density takes the form
\be
\delta \epsilon(n,h) = {m_h^2 v^2 \over 2} \, \chi^2 \left (1+{\chi \over 2} \right )^2 +  {(3 \pi^2 \, n)^{5/3} \over 10 \pi^2 M_2}+ {m^2 \, n^2 \over \Delta^2 \, v^2 } - {m^2 \over \Delta} \, \chi \left (1+{\chi \over 2} \right ) \, n~.
\label{ded}
\ee
It can be readily verified that the necessary condition for this expression to develop a negative minimum is $m > \sqrt{2} m_h \approx 177$\,GeV, which would be problematic from the point of vacuum stability~\cite{dveegis}.

Thus only the case $\Delta \ll M$ appears to be interesting with regards to the existence of the condensate. In this case for a sufficiently small $\Delta$ one can find $m (1+\chi) \gg \Delta$ so that the effect of the mass splitting $\Delta$ is small. Using the formulas in Eq.(\ref{chim}) and Eq.(\ref{nmin}) it can be estimated that this regime is achieved if $\Delta/M$ is much smaller than $(m_h/v)^{4/3}$, if $m$ and $m_h$ are considered as being of the same order.

The presented estimates are based on considering the energy density of a spatially uniform condensate of the heavy neutrinos. Such treatment  is justified only inasmuch as the surface energy can be neglected in comparison with the bulk energy, associated with the transition of the Higgs field on the surface of the globule, and requires the radius of the globule $R$ to be large. Considering the parameters $m$ and $m_h$ as being of the same order and describing the smallest mass scale in the problem, one can conservatively estimate that the condition for the approximation used is  $R \, m_h \gg 1$. According to the estimate in Eq.(\ref{nmin}) for the equilibrium number density at large $M$, the total number ${\cal N}$ of particles in the globule should then satisfy the condition
\be
{\cal N} \gg { 4 (8 \, r \, m/m_h - 2 - r^2)^3 \over 729 \, \pi^3 r^2} \, {  M^3 \over m_h \, v^2}~.
\label{npart}
\ee

The estimate (\ref{npart}) of the total number of particles naturally places a limit on how big the mass $M$ can be before the globule collapses into a Black Hole. A straightforward estimate then gives the condition
\be
M \ll \sqrt{ M_{\rm Pl} v} \sim 10^{11}\, {\rm GeV}~,
\label{bhb}
\ee 
still leaving an ample range of the values of $M$ at which the discussed globules may exist. It is also interesting to notice that within this range the scale for the mass $M {\cal N}$ of the globule can reach macroscopic values: 
$$M^4/(m_h v^2) \approx \left ( {M \over 10^8 \, {\rm GeV}} \right )^4 \, 20\,{\rm g}~,$$
while the scale for its radius is set by $1/m_h \sim 10^{-16}$\,cm. 

The possibility that the discussed here globules do indeed exist makes it very interesting to study their phenomenological implications, in particular their interaction with ordinary matter. Any detailed discussion of this subject is beyond the scope of the present paper. However, it can be mentioned, as a simple remark, that the enhanced Higgs field inside the globule results in a strong repulsion from the ordinary matter. Indeed, the masses of leptons and quarks are increased inside the condensate by the factor $(1+\chi)$ and the QCD mass scale $\Lambda_{QCD}$ is larger inside the globule due to the larger masses of quarks and the resulting cutoff at a higher scale of the screening contribution to the running of the QCD coupling. For this reason it is impossible to exclude existence of the globules from the apparent absence of such objects bound in ordinary matter.

In summary. In an extension of the Standard Model with vectorlike leptons it is found that in a certain part of the parameter space the lighter (but still superheavy) neutral fermions can assemble into globules with the number density $n$ exceeding the EW scale $m_h v^2$. The energy per particle inside the globule is lower than the fermion mass in the vacuum due to the strong enhancement of the Higgs field inside the condensate by the collective effect of the Yukawa interaction. The mass of a globule can reach a macroscopic scale (i.e. up to grams or more), and the globules are absolutely stable if the constituent neutral fermions are stable and can be a part or all of the dark matter. However, if at present the vectorlike fermions exist in a form of dispersed individual particles, their spontaneous condensation into globules is prevented by an energy barrier. If the globules could still be produced during the cosmological evolution, a study of their properties, such as interaction with ordinary matter, can result in new limits on models with vectorlike fermions, or in a realization of a possible new form of the dark matter.

This work  is supported, in part, by U.S. Department of Energy Grant No. DE-SC0011842.

\end{document}